\documentstyle[epsfig,url,times]{aa}

\def\B{BeppoSAX}
\def\S{XTE J0421+560}
\title{X--ray/optical observations of \S/CI Cam in quiescence}

\author{M.~Orlandini\inst{1}
\and A.N.~Parmar\inst{2}
\and F.~Frontera\inst{1,3}
\and N.~Masetti\inst{1}
\and D.~Dal~Fiume\inst{1}
\and A.~Orr\inst{4}
\and A.~Piccioni\inst{5}
\and G.~Raimondo\inst{6}
\and A.~Santangelo\inst{7}
\and G.~Valentini\inst{6}
\and T.~Belloni\inst{8}
}

\institute{
Istituto Tecnologie e Studio Radiazioni Extraterrestri, TeSRE/CNR, Via
  Gobetti 101, I--40129 Bologna, Italy
\and
Astrophysics Division, Space Science Department of ESA, ESTEC, NL--2200
  AG Noordwijk, The Netherlands
\and
Dipartimento di Fisica, Universit\`a di Ferrara, Via Paradiso 11,
  I--44100 Ferrara, Italy
\and
Institute for Astronomy and Astrophysics T\"ubingen, IAAT, Waldhaeuser
  Stra{\ss}e 64, D--72076 T\"ubingen, Germany
\and
Dipartimento di Fisica, Universit\`a di Bologna, Via Zamboni 33,
  I--40126 Bologna, Italy
\and
Osservatorio Astronomico di Teramo, Collurania, I-64100 Teramo, Italy
\and
Istituto Fisica Cosmica e Applicazioni all'Informatica, IFCAI/CNR, via
  La Malfa 153, 90146 Palermo, Italy
\and
Osservatorio Astronomico di Brera, Via Bianchi 46, I--23807 Merate (Lc),
  Italy
}

\offprints{orlandini@tesre.bo.cnr.it}
\date{Received \today; Accepted }
\thesaurus{06          
           (08.02.5;   
            08.09.2;   
            08.14.2;   
            13.25.1;   
            13.25.5)   
}

\begin{document}

\maketitle

\markboth{M.\ Orlandini et al.: X--ray/optical observations of \S/CI Cam
in quiescence}{M.\ Orlandini et al.: X--ray/optical observations of
\S/CI Cam in quiescence}

\begin{abstract}

We report on a \B\ observation of the transient X--ray source
\object{\S} during quiescence performed $\sim$150~days after the 1998
April outburst. The source had an unabsorbed 0.5--10~keV flux of
$6.7\times 10^{-12}$~erg~cm$^{-2}$~s$^{-1}$ and was still remarkably
soft with most of the emission below 2~keV. The X--ray spectrum can be
fit with the same two-temperature model as the outburst observations.
There is evidence for the presence of an Iron emission feature at
$\sim$7~keV. We report also on a series of optical observations
performed using the 72~cm Teramo-Normale Telescope (TNT) of the Teramo
Observatory, and the 1.5~m Loiano Telescope of the Bologna Observatory.
The optical spectrum includes very strong Balmer emission lines, He~I,
[Fe~II], and [O~I] features. From the observed $L_x/L_{\rm opt}\sim
10^{-3}$ the quiescent optical emission cannot be due to re-processing
of the X--rays, but has to be generated in the optical companion or the
circumstellar material. Moreover, the quiescent X--ray luminosity cannot
be due to the optical star if it is of spectral type OB. Although the
nature of the compact object present in the \S/CI Cam system cannot be
firmly established,  we speculate that it is most probably a white
dwarf.


\end{abstract}

\section{Introduction}

The soft X--ray transient (SXT) \S\ was the target of a series of
observations by CGRO (\cite{1845}), RXTE (\cite{1836}), ASCA
(\cite{1790}) and \B\ (\cite{1730}) soon after an outburst discovered by
the All-Sky Monitor (ASM) on-board RXTE in 1998 April (\cite{1834}). 
Radio observations of \S\ revealed the presence of radio jets
(\cite{1843}), similar to those observed from \object{SS433}. The jet
motion was estimated to be $\sim$26~mas/day, corresponding to a velocity
of $0.15~c$ assuming a source distance of 1~kpc (\cite{1739}).
Subsequent radio observations have revealed a slow
($\sim$1000~Km~s$^{-1}$), shell-like motion (\cite{1847}), confirmed by
optical observations (\cite{2032}).

The optical counterpart of \S\ is the peculiar star \object{CI Cam} (aka
\object{MWC 84}) which, on the basis of the new classification criteria
proposed by Lamers et~al.\ (1998), \nocite{1872} is classified as a B[e]
star. This classification includes supergiants, pre-main sequence stars,
compact planetary nebulae, symbiotic stars and unclassified stars.
Lamers et~al.\ (1998) include CI Cam among the unclassified stars, while
Zorec (1998) \nocite{1879} includes it among the proto-planetary nebulae
with dusty circumstellar envelopes. The binary nature of CI Cam has been
deduced from the extremely large ratio between IRAS 84~$\mu$m and 
12~$\mu$m fluxes, which is close to that of binary systems that contain
both a hot and a cool star (\cite{1750}).

We have already reported on the two Target of Opportunity (TOO)
observations performed by \B\ during the 1998 April outburst
(\cite{1774}; \cite{1775}). The energy spectra cannot be fit by any
simple model, and displayed a dramatic change from TOO1 (performed on
1998 April 3) to TOO2  (1998 April 9), namely the onset of a soft
(E$\la$2 keV) component (\cite{1775}). A two-temperature bremsstrahlung
model was used to describe the spectra of both TOOs, with temperatures
($kT_1$, $kT_2$)$\sim$(1.27, 6.81) keV for TOO1, and $\sim$(0.20, 2.78)
keV for TOO2 (\cite{1775}). The spectra for both TOOs included line
features, identified with O, Ne/Fe-L, Si, S, Ca and Fe-K. During TOO2
the O and Ne/Fe-L line energies decreased smoothly by $\sim$9\%, while
the other lines did not show any shifts. Because of the peculiar
temporal variability of the source, Frontera et~al.\ (1998)
\nocite{1774} associated the soft component to the relativistic jets,
and the hard component to processes occurring in circumstellar matter.

The nature of the compact object responsible for the X--ray emission is
still unknown and controversial. The low energetics involved in the
outburst and the absence of erratic time variability are unlike
outbursts from neutron star or black hole X--ray novae (\cite{1774}).
However the presence of X--ray, optical and radio emission, together
with relativistic jets, is typical of neutron star and black hole
systems.

In order to try to elucidate the nature of the compact object by
studying the source spectral evolution during its quiescent phase, we
performed a third TOO observation of \S\ as soon as the source was once
again observable by \B. In the next section we detail the data analysis
performed on the X--ray data, while in Sect.~3 we present results from
optical observations performed at the Teramo and Bologna astronomical
observatories. Finally, in Sect.~4 we discuss the implications of these
observations.

\section{X--ray Observation and data analysis}

The \B\ satellite (\cite{1530}) is equipped with four narrow field
instruments (NFIs) able to cover the unprecedented wide 0.1--200 keV
energy band. Two NFIs are imaging instruments, namely the LECS
(\cite{1531}) and MECS (\cite{1532}), operating in the 0.1--10~keV and
1.5--10~keV energy bands, respectively.  The other two NFIs are
mechanical collimated instruments: the  HPGSPC (\cite{1533}) and PDS
(\cite{1386}). The former operates in the 3--180~keV band, while the
latter operates in the 15--200~keV band. The background for the
non-imaging NFIs is monitored by rocking the collimators off-source by
$3^\circ$ (one-side rocking) for the HPGSPC, and 3\fdg5 (two-side
rocking) for the PDS.

The third \B\ observation of \S\ was performed between 1998 September 3
14:19 UT and September 4 13:47. This is 157 days after the \S\ burst
peak of 1998 April 1 (the long delay between TOO2 and TOO3 is due to
unobservability of the source  with \B\ because of viewing constraints).
All four NFIs worked nominally during the observation. Data were
collected in direct mode, providing information on time, energy and
position (for the imaging instruments), and were processed using the
{\sc saxdas} 2.0 package (\cite{1650}) except for PDS data, where {\sc
xas} 2.1 (\cite{1619}) was used. Good data were selected using default
criteria. For the imaging instruments, data were extracted from circular
regions centered on the source position, with a $4'$ radius for the
LECS, and $2'$ for the MECS (the smaller MECS extraction region is due
to S/N optimization above 4 keV).

Because of the low galactic latitude of the source, we did not use the 
standard background LECS blank field measurement, but instead the
background was estimated from two semi-annuli near the outside of the
field of view (\cite{1832}). The 3$\sigma$ uncertainty obtained using
this new method is $\la$2.2$\times 10^{-3}$ c~s$^{-1}$ (0.1--10~keV), or
$\la$3.7$\times 10^{-3}$ c~s$^{-1}$ (0.1--2 keV). To estimate the MECS
background we used both a standard file and a background estimated from
an annular region centered on the source position with inner and outer
radii of $4'$ and $8'$, respectively. The results do not depend
significantly on which MECS background was used.  For the collimated
instruments, we evaluated the background from the offset fields, using a
standard background subtraction procedure. The offset fields were also
checked for the presence of contaminating sources.

\begin{figure*}
\epsfysize=7cm 
\centerline{\epsffile{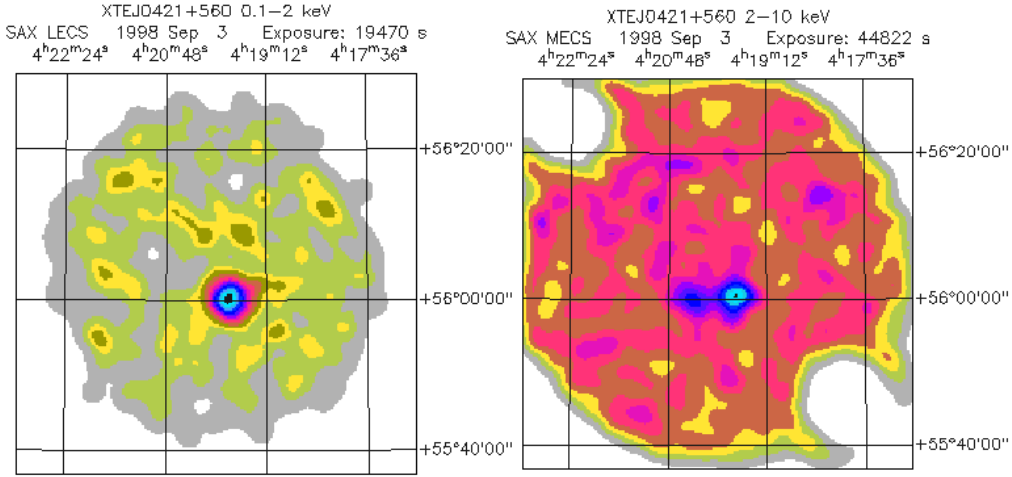}}
\caption[]{LECS (left) and MECS (right) $4\sigma$ Gaussian smoothed
images of the \S\ field obtained during the TOO3 \B\ observation}
\label{fig:lecs_mecs_img}
\end{figure*}

A faint source at a position consistent with \S\ was detected in both
imaging instruments (see Fig.~\ref{fig:lecs_mecs_img}). The 0.1--10~keV
LECS net count rate is $0.0037\pm 0.0006$ c~s$^{-1}$, while the
1.5--10~keV MECS count rate is $0.0024\pm 0.0003$ c~s$^{-1}$. The
exposure times for the  LECS and MECS are 19~ks and 45~ks, respectively. 
This difference is due to the constraint that the LECS can only be
operated in spacecraft night time. The 2$\sigma$ PDS upper limit is 0.5
mCrab (15--30~keV).

\begin{figure}
\centerline{\epsfig{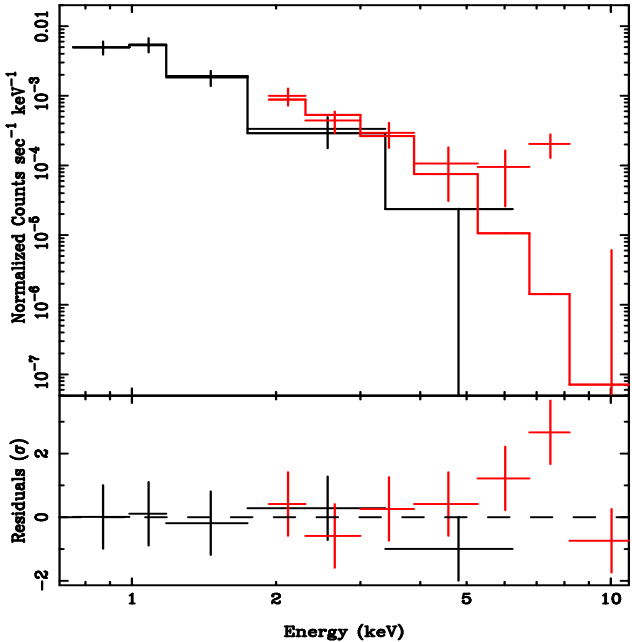}}
\caption[]{LECS/MECS count rate spectra {\em (plus signs)} and 2BREMS
best fitting continuum plus an Iron line {\em (histogram)}, together
with the fit residuals. The Gaussian normalization has been set to zero,
in order to show its profile in the residual panel}
\label{fig:lecs_mecs_spc}
\end{figure}

The combined LECS and MECS spectrum can be fit with the same continuum
model as  used for TOO1 and TOO2, namely a two-temperature
bremsstrahlung (2BREMS) model. A factor was included in the spectral
fitting to allow for known normalization differences between the two
instruments (\cite{1827}). The fit yields a $\chi^2$ of 10.9 for 6
degrees of freedom (dof). The inclusion of a narrow Gaussian emission
line at $\sim$7~keV improved the fit, yielding a $\chi^2$/dof of 2.5/4.
The inclusion of this line is marginally significant, the probability of
chance improvement (PCI) of the $\chi^2$, computed by means of an
F-test, being equal to 5.3\%. In Fig.~\ref{fig:lecs_mecs_spc} the 2BREMS
plus Iron line fit to the LECS and MECS spectra is shown, together with
the fit residuals. The line normalization has been set to zero in the
lower panel to display its profile. In Table~\ref{tab:fits} the best fit
parameters for the 2BREMS model, together with other continuum models
used to fit the data: a simple power-law and a one-temperature
bremsstrahlung are presented. A simple black-body spectrum does not fit
our data, with a $\chi^2$/dof of 17.6/7. The amount of photo-electric
absorption was not well constrained with any of the  models, and it was
therefore fixed at the galactic value (\cite{1616}). The unacceptable
fit using the 1BREMS model is due to the need for a second component
above 5 keV. The PCI of the $\chi^2$ from the PL to the 2BREMS model is
56\%. We also tried to fit the spectrum using a two-temperature
equilibrium plasma emission model (\cite{1855}), as performed with ASCA
data (\cite{1790}). The fit yields a $\chi^2$/dof of 9.6/7, with large
positive residuals near $\sim$7~keV.

\def\mcc#1{\multicolumn{1}{c}{#1}}
\def\er#1#2{_{-#1}^{+#2}}
\begin{table}
\caption[]{Best fit spectral parameters. PL: Power-law; 1BREMS:
one-temperature bremsstrahlung; 2BREMS: two-temperature bremsstrahlung.
We also list the Probability of Chance Improvement (PCI) of the $\chi^2$
when a Gaussian emission line is added, computed from an F-test.
Uncertainties are given at 90\% confidence level for a single parameter}
\label{tab:fits}
\begin{flushleft}
\begin{tabular}{llll}
\hline\noalign{\smallskip}
\mcc{Parameter}  & \multicolumn{3}{c}{Value} \\ 
\noalign{\smallskip} \cline{2-4} \noalign{\smallskip}
 & \mcc{PL} & \mcc{1BREMS} & \mcc{2BREMS} \\
\noalign{\smallskip} \hline \noalign{\smallskip}
LECS/MECS               & $0.6\er{0.3}{0.4}$ & $0.4\er{0.3}{0.4}$    & $0.6\er{0.5}{0.7}$    \\
N$_{\rm H}$ ($10^{22}$ cm$^{-2}$) & $0.495^a$ & $0.495^a$             & $0.495^a$            \\
$\alpha$                & $5.3\er{0.8}{0.9}$ &                       &                       \\
kT$_1$ (keV)            &                    & $0.35\er{0.11}{0.07}$ & $0.22\er{0.09}{0.08}$ \\
kT$_2$ (keV)            &                    &                       & $1.3\er{0.7}{6.7}$    \\
E$_{\rm Fe}$ (keV)      & $7.0\er{0.2}{1.0}$ & $6.9\er{0.2}{1.0}$    & $7.0\er{0.2}{1.0}$    \\
$\sigma_{\rm Fe}$ (keV) & 0.1$^b$            & 0.1$^b$               & 0.1$^b$               \\
I$_{\rm Fe}\ \ ^c$      & $6.7\er{3.9}{3.5}$ & $7.7\er{4.7}{3.0}$    & $6.2\pm 3.8$          \\
$\chi^2$/dof            & 5.48/7             & 14.24/7               & 2.80/5                \\
PCI                     & 0.0512             & 0.22                  & 0.0126                \\
\noalign{\smallskip} \hline \noalign{\smallskip}
\multicolumn{4}{l}{$^a$ Fixed at the galactic value (\cite{1616})} \\
\multicolumn{4}{l}{$^b$ Fixed at the MECS resolution} \\
\multicolumn{4}{l}{$^c$ In units of 10$^{-6}$ Photons cm$^{-2}$ s$^{-1}$}
\end{tabular}
\end{flushleft}
\end{table}

The observed 0.5--2~keV and 2--10~keV fluxes computed from the 2BREMS
model are $7.5\times 10^{-13}$ and $1.5\times 10^{-13}$
erg~cm$^{-2}$~s$^{-1}$, respectively. The unabsorbed fluxes (computed by
setting \mbox{N$_{\rm H}$}$\equiv$0) in the 0.5--2~keV and 2--10~keV
energy ranges are $6.5\times 10^{-12}$ and $1.7\times 10^{-13}$
erg~cm$^{-2}$~s$^{-1}$, respectively. Using these values, we obtain
X--ray luminosities of $7.8\times 10^{32}$ (0.5--2 keV) and $2.0\times
10^{31}$ (2--10 keV) d$^2_{\rm kpc}$ erg~s$^{-1}$, where d$_{\rm kpc}$
is the distance to \S\ in kpc. These estimates are strongly affected by
the uncertainty in the N$_{\rm H}$ value: a 50\% increase in N$_{\rm H}$
corresponds to a 60\% increase in the 0.5--2 keV luminosity. On the
other hand $kT_1$ is insensitive to N$_{\rm H}$, with only a 18\%
decrease in the best fit temperature with respect to a 50\% increase in
N$_{\rm H}$.

\section{Optical observations and data analysis}

The field of CI Cam was imaged on 1998 September 3, simultaneously with
the \B\ observation, with the 72~cm Teramo-Normale Telescope (TNT) of
the Teramo Observatory. A total of 15 frames in the $B$, $V$, $R$ and
$I$ bands were acquired between September 3.980 and September 4.048, 
with exposure times between 1 and 15 minutes. The frames were corrected
for bias and flat fielded in the standard fashion and then reduced with
the {\sc daophot ii} package (\cite{1852}) and the PSF-fitting algorithm
{\sc allstar} inside {\sc midas}. The star, during these observations,
had $V = 11.83\pm 0.05$, $B$--$V = 0.81\pm 0.07$, $V$--$R = 0.82\pm
0.07$ and $V$--$I = 1.58\pm 0.07$, with no substantial luminosity
variations amongst the frames acquired in each single filter.

Spectroscopy on this object was then performed on 1998 December 14 and
1999 January 23 with the 1.5~m telescope of the Bologna Astronomical
Observatory. Spectra with grisms \#3 (3000--6500 \AA), \#4 (4000--9000
\AA) and \#5 (5000--10000 \AA) were acquired with a slit width of
$2\farcs5$, which gave dispersions of 2.1, 3.0 and 2.8 \AA/pixel,
respectively. The exposure times ranged from 2 to 30 minutes.
Table~\ref{tab:opt_log} is a log of the spectrophotometric observations.

\begin{figure*}
\centerline{\epsfig{file=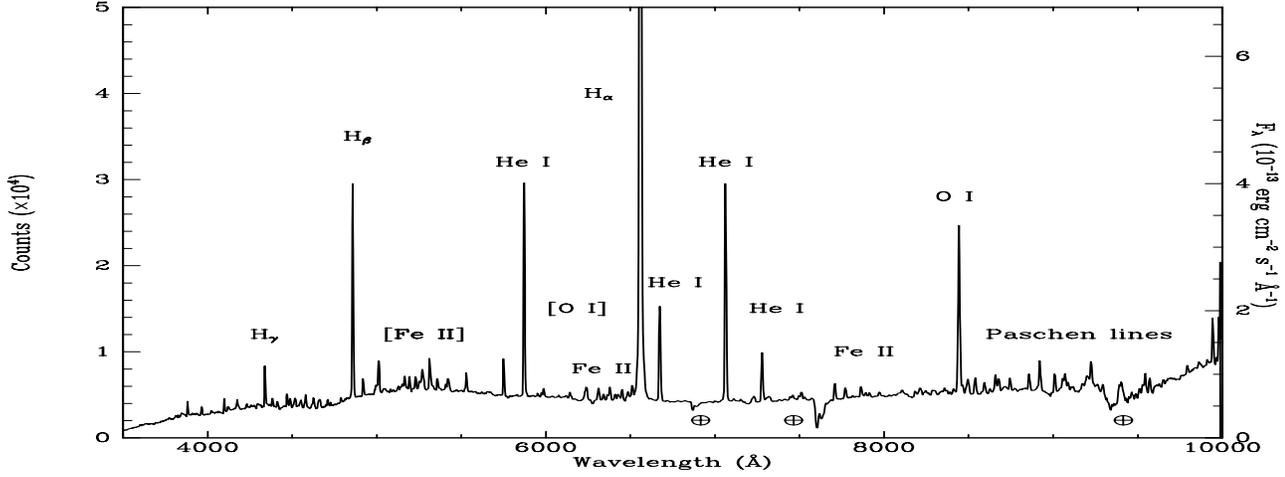,height=.8\textwidth,width=10cm,angle=270}}
\vspace*{-3.5cm}
\caption[]{The optical spectrum of CI Cam taken at the 1.5~m telescope
of the Bologna Astronomical Observatory on 1999 January 23. Telluric
absorption features are marked with $\oplus$}
\label{fig:opt_spc}
\end{figure*}

\def\mcl#1{\multicolumn{1}{l}{#1}}
\begin{table}
\caption[]{Log of spectrophotometric observations on CI Cam. Imaging
(upper part) and spectra (lower part) sequences are reported}
\label{tab:opt_log}
\begin{flushleft}
\begin{tabular}{cccc}
\hline\noalign{\smallskip}
Date & Start time & Filter or & Exposure \\
     & (UT) & passband & (minutes) \\
\noalign{\smallskip} \hline\noalign{\smallskip}
\multicolumn{4}{c}{Imaging} \\
\noalign{\smallskip} \hline\noalign{\smallskip}
\mcl{1998 Sep 3} & 23:30:33 & $I$ & 1 \\
\mcl{1998 Sep 3} & 23:35:01 & $I$ & 1 \\
\mcl{1998 Sep 3} & 23:37:41 & $I$ & 1 \\
\mcl{1998 Sep 3} & 23:39:19 & $R$ & 1 \\
\mcl{1998 Sep 3} & 23:40:51 & $R$ & 2 \\
\mcl{1998 Sep 3} & 23:43:27 & $R$ & 2 \\
\mcl{1998 Sep 3} & 23:45:40 & $R$ & 2 \\
\mcl{1998 Sep 3} & 23:48:40 & $V$ & 3 \\
\mcl{1998 Sep 3} & 23:52:30 & $V$ & 4 \\
\mcl{1998 Sep 3} & 23:56:45 & $V$ & 4 \\
\mcl{1998 Sep 4} & 00:01:24 & $B$ & 4 \\
\mcl{1998 Sep 4} & 00:06:13 & $B$ & 10 \\
\mcl{1998 Sep 4} & 00:17:42 & $B$ & 10 \\
\mcl{1998 Sep 4} & 00:28:38 & $B$ & 15 \\
\mcl{1998 Sep 4} & 00:45:58 & $B$ & 15 \\
\mcl{1998 Sep 4} & 01:03:16 & $B$ & 15 \\
\\
\noalign{\smallskip} \hline\noalign{\smallskip}
\multicolumn{4}{c}{Spectra} \\
\noalign{\smallskip} \hline\noalign{\smallskip}
\mcl{1998 Dec 14} & 22:07:35 & Grism \#4 & 30 \\
\mcl{1998 Dec 14} & 22:41:05 & Grism \#4 & 10 \\
\mcl{1998 Dec 14} & 22:54:07 & Grism \#4 & 3 \\
\mcl{1998 Jan 23} & 19:17:31 & Grism \#3 & 30 \\
\mcl{1998 Jan 23} & 19:52:46 & Grism \#3 & 30 \\
\mcl{1998 Jan 23} & 20:25:49 & Grism \#4 & 2 \\
\mcl{1998 Jan 23} & 20:30:58 & Grism \#4 & 8 \\
\mcl{1998 Jan 23} & 20:42:01 & Grism \#4 & 8 \\
\mcl{1998 Jan 23} & 20:54:11 & Grism \#5 & 8 \\
\mcl{1998 Jan 23} & 21:04:49 & Grism \#5 & 8 \\
\mcl{1998 Jan 23} & 21:16:10 & Grism \#5 & 2 \\
\noalign{\smallskip} \hline
\end{tabular}
\end{flushleft}
\end{table}

Spectra were debiased and flat-fielded in the standard way and then
extracted and reduced with {\sc iraf}. Wavelength calibration was made
using He--Ar lamps. Flux calibration was performed only for the December
14 spectra using the spectroscopic standard HD~60778. The January 23
spectra were not flux calibrated because of poor atmospheric conditions.
Due to the constancy, within the uncertainties, of the $V$ band
magnitude between the two observations as shown by the VSOLJ data
(\cite{1849}), and because of non variability of the main emission line
ratios in the two data sets, we assumed that the flux level of the
continuum remained constant during the two runs, and so we calibrated
the January 23 spectra with the same standard as December 14.

The summed (3500--10000 \AA) optical spectrum of CI Cam for the night of 
January 23 is shown in Fig.~\ref{fig:opt_spc}. It reveals very strong
emission lines of the Balmer series of Hydrogen, plus He~I, [Fe~II],
[O~I] and weak He~II (see Table~\ref{tab:opt_lines}). A quick-look
comparison with the spectrum of Downes (1984) \nocite{1823} acquired 15
years before shows that all the main emission lines identified by that
author are still present; moreover, in the red branch --- which was not
covered by Downes --- a strong O~I line at $\lambda8446$ and the Paschen
series in emission are seen. We note that the strength of the He~I lines
is strongly reduced. The He~II emission lines at $\lambda\lambda4686$
and 5411 are marginally detected. In Table~\ref{tab:opt_lines} we detail
some of the lines detected in the CI Cam spectrum, together with their
equivalent widths and line fluxes.

We de-reddened the optical data using the N$_{\rm H}$ column value
computed in the direction of CI Cam (N$_{\rm H} = 4.95\times 10^{21}$
cm$^{-2}$; Daltabuit \& Meyer 1972). We then fit the 4000--10000 \AA\
optical spectrum with a power-law ($F(\lambda) \propto \lambda^\alpha$),
yielding a power-law index of $-2.43\pm 0.01$. The total optical
unabsorbed flux is $4.0\times 10^{-9}$ erg~cm$^{-2}$~s$^{-1}$,
corresponding to an optical luminosity $L_{\rm opt} = 4.8\times 10^{35}$
d$^2_{\rm kpc}$ erg~s$^{-1}$.

The mean values reported in Table~\ref{tab:opt_lines} for the EWs of the
[Fe~II] $\lambda$4414 and [O~I] $\lambda$6364 forbidden emission lines
were computed using only the January spectra. This is due to the low S/N
ratio for these features in the December observations. Since the EWs of
[Fe~II] $\lambda\lambda$5527 and 5746 lines, as well as those of the
permitted emission lines, did not vary substantially between the two
spectroscopic runs, it is likely that also the EWs of [Fe~II]
$\lambda$4414 and [O~I] $\lambda$6364 remained constant within the
errors.

It is noteworthy that, for at least the [Fe~II] $\lambda\lambda$5527 and
5746, a shift of about 8~\AA\ is observed with respect to the main
permitted emission lines such as the Balmer series, the He~I, O~I and
Fe~II lines. A similar shift cannot be verified for [Fe~II]
$\lambda$4414 and [O~I] $\lambda$6364, because of their poorer S/N ratio
and the insufficient spectral resolution, especially for the [O~I] line.
This suggests different origins and dynamics for the permitted and at
least some of the forbidden lines. We also note the presence of weak
interstellar absorption bands at $\lambda$5780 (EW $=0.4\pm 0.1$) and
$\lambda$6284 (EW $=1.6\pm 0.3$).

\begin{table}
\caption[]{Absolute values of the mean equivalent widths (expressed in
\AA ngstroms) and line fluxes of the main emission lines identified in
the spectrum of CI Cam}
\label{tab:opt_lines}
\begin{flushleft}
\begin{tabular}{l@{\ }lr@{$\,\pm\,$}lll}
\hline\noalign{\smallskip}
\multicolumn{2}{c}{Line} & 
\multicolumn{2}{c}{EW (\AA)}    & \mcc{Line Flux$^a$} & \mcc{$\log L_{\rm line}$\ $^b$} \\
\noalign{\smallskip} \hline\noalign{\smallskip}
H$_\alpha$  &               & 414.2& 49.5 & $(2.3\pm 0.2)\times 10^{-10}$ & 34.44 \\
H$_\beta$   &               & 48.0 & 5.4  & $(7.7\pm 0.8)\times 10^{-11}$ & 33.96 \\
H$_\gamma$  &               & 10.4 & 1.2  & $(2.6\pm 0.4)\times 10^{-11}$ & 33.49 \\
He~I        & $\lambda$4471 & 3.6  & 0.8  & $(7.2\pm 1.4)\times 10^{-12}$ & 32.93 \\
            & $\lambda$5875 & 46.1 & 4.4  & $(2.8\pm 0.3)\times 10^{-11}$ & 33.52 \\
            & $\lambda$6678 & 29.8 & 3.2  & $(1.4\pm 0.1)\times 10^{-11}$ & 33.22 \\
            & $\lambda$7165 & 69.5 & 7.8  & $(2.8\pm 0.3)\times 10^{-11}$ & 33.52 \\
            & $\lambda$7281 & 15.9 & 1.7  & $(5.7\pm 0.6)\times 10^{-12}$ & 32.83 \\
He~II       & $\lambda$4686 & 0.25 & 0.1  & $(4.8\pm 1.9)\times 10^{-13}$ & 31.76 \\
            & $\lambda$5411 & 0.8  & 0.2  & $(6.6\pm 1.7)\times 10^{-13}$ & 31.90 \\
O~I         & $\lambda$8446 & 49.4 & 5.2  & $(1.8\pm 0.2)\times 10^{-11}$ & 33.33 \\
\noalign{\smallskip} \hline\noalign{\smallskip}
$[$Fe~II$]$ & $\lambda$4414 & 1.6 & 0.4   & $(3.2\pm 0.7)\times 10^{-12}$ & 32.58 \\
$[$Fe~II$]$ & $\lambda$5527 & 4.2 & 0.5   & $(3.5\pm 0.5)\times 10^{-12}$ & 32.62 \\
$[$Fe~II$]$ & $\lambda$5746 & 8.8 & 1.0   & $(6.2\pm 0.9)\times 10^{-12}$ & 32.87 \\
$[$O~I$]$   & $\lambda$6364 & 1.6 & 0.3   & $(8.5\pm 1.7)\times 10^{-13}$ & 32.01 \\
\noalign{\smallskip} \hline\noalign{\smallskip}
\multicolumn{6}{l}{$^a$ In units of erg~cm$^2$~s$^{-1}$} \\
\multicolumn{6}{l}{$^b$ In units of d$^2_{\rm kpc}$ erg~s$^{-1}$}
\end{tabular}
\end{flushleft}
\end{table}

Using the ratios of some He~I emission lines, particularly
$\lambda\lambda$4471, 5875 and 6678, it is possible to evaluate the
$E(B$--$V)$ color excess (see Della Valle \& D\"urbeck (1993)
\nocite{1848} and references therein). Comparing the theoretical and
observed ratios of $\lambda\lambda$5876/4471 and
$\lambda\lambda$6678/4471 we derive $E(B$--$V) = 1.47$ and 1.60,
respectively. This gives a mean $E(B$--$V) = 1.54$. Alternatively, the
H$_\alpha$/H$_\beta$ line ratio yields the lower value $E(B$--$V) =
1.02$; we anyway caution that in some cases the Balmer decrement may not
be attributed only to the reddening effect (\cite{1853}). Nevertheless,
using the mean of the values derived from the H and He line ratios we
obtain $E(B$--$V) = 1.28$. This corresponds to a $V$ band absorption
$A_V = 4.0$ and, using the empirical formula of Predehl \& Schmitt
(1995), \nocite{1851} to an hydrogen column density $N_{\rm H} =
7.1\times 10^{21}$ cm$^{-2}$. This value is about 40\% higher than the
Galactic hydrogen column density in the direction of CI Cam
(\cite{1616}). This suggests that intrinsic absorption plays an
important role. Indeed, from our measured value of N$_H$, consistent
with that measured during TOO2, we derive a 90\% confidence interval of
4.13--$6.03\times 10^{21}$ cm$^{-2}$ which translates to $2.3<A_V<3.4$.
Belloni et~al.\ (1999) \nocite{1873} obtained a value of A$_V = 4.36$ by
fitting the optical/IR spectrum of CI Cam with a combined Kurucz plus
optical thin dust model (\cite{1883}). Alternatively, Zorec (1998)
derived for this source the higher value of A$_V = 4.88$, which he
interpreted as partly (2.88) due to the interstellar medium and partly
(2.40) due to a dusty circumstellar envelope.

Doppler analysis of the most prominent emission lines of the two sets of
spectral data shows that no substantial line shifts are detectable
within the same night, {\em i.e.\/} the centroid line variations fall
inside the measurement uncertainty of 0.1 \AA. The same result is found
when the December and January data sets are compared. Because it is
quite likely that most of the observed lines originate in the ionized
shell shrouding the system, the absence of line shifts cannot be used to
establish a firm upper limit on the projected orbital velocity.

\section{Discussion}

The major problem in understanding the \S/CI Cam system is the fact that
the nature of the components in the system is not well established.
Miroshnichenko (1995), by fitting the optical continuum, derived a
spectral type for the two CI Cam components as B0~V and K0~II, and an
interstellar extinction A$_V$ of 1.1. The derived spectral type implies
a distance of 6~kpc to reconcile with the observed $V$ magnitude.
Miroshnichenko (1995) also fitted the optical continuum with a single
star model, obtaining the same spectral type for the hot star, but a
higher extinction (A$_V=3.7$). In the latter case, a distance of 2.4~kpc
was estimated. With the value of $A_V = 4.88$ estimated by Zorec (1998)
a distance of 1.75~kpc was instead derived. From an observation of the
H~I absorption profile at 21~cm, Hjellming (private communication)
inferred a distance of CI Cam of $1.0 \pm 0.2$ kpc, that is smaller than
that  inferred from the Zorec calculations, but not inconsistent with
the measured $V$ magnitude, if an extinction A$_V$ of the order of 4 is
assumed. It is therefore likely that the normal object is a hot star
showing the B[e] phenomenon, with a dusty envelope responsible for the
IR emission. This is confirmed by optical observations showing a
spherical-symmetric shell expanding at $\sim$32~Km~s$^{-1}$ and present
before the X--ray outburst, and an asymmetric shell, moving at
$\sim$2000~Km~s$^{-1}$, which emerged from the low-velocity shell soon
after the outburst (\cite{2032}). From the observed $L_x/L_{\rm opt}\sim
10^{-3}$ the quiescent optical emission cannot be due to re-processing
of the X--rays, but has to be generated in the optical companion or the
circumstellar material.

The nature of the compact object present in \S\ is not revealed by the
observation of X--ray phenomena typical of black hole (BH), neutron star
(NS), or white dwarf (WD) systems, therefore their presence can only be
inferred indirectly by comparison with the phenomenology observed in
other systems in which the compact object is known. The weakness of this
approach is the peculiarity of the \S\ phenomenology, that makes it a
unique system amongst the class of SXTs.

In the presence of a BH system an X--ray outburst can occur either via
an accretion disc instability or through a mass transfer instability
(see e.g.\ \cite{1871}). The latter instability is unlikely to work in
\S\ because of the lack of hard ($\ga$7 keV) X--ray photons in the
quiescent spectrum. Indeed, if the transient event is due to mass loss
instability that arises in the secondary star as a result of X--ray
illumination of the atmosphere, then hard X--ray photons are required in
order to start the instability and therefore produce the outburst
(\cite{1860}). The observed quiescent X--ray luminosity is not
sufficient to trigger this mass-overflow instability (see e.g.\
\cite{2035}). Indeed, for this instability to work it is necessary that
the X--ray flux (E$\ga$7 keV) at the inner Lagrangian point $L1$ and the
stellar flux at $L1$ be comparable (\cite{1860}). The X--ray flux at
$L1$ (assuming isotropic emission) is $L_x/4\pi b^2$, where $b$ is the
distance between $L1$ and the compact object. The stellar flux at $L1$
is $L_{\rm opt}/4\pi (a-b)^2$, where $a$ is the orbital separation.
Therefore the mass-transfer instability will work if $L_x/L_{\rm opt}\ga
(b/(a-b))^2\approx (M_x/M_{\rm opt})^2$, where $M_x$ and $M_{\rm opt}$
are the compact object and the companion masses, respectively. From our
observed value $L_x(\ga$7~kev)/$L_{\rm opt}\sim 10^{-5}$ we conclude
that it is quite unlikely for the mass-overflow instability to take
place in the \S/CI Cam system. The presence of an accretion disc would
produce a strong temporal variability, and double peaked emission lines
in the optical spectrum of SXTs, expecially H$_\alpha$ and H$_\beta$
(\cite{1857}; \cite{1859}). While temporal variability was observed in
\S\ only in the soft (E$<$1~keV) energy range soon after the burst and
attributed to the relativistic jets (\cite{1774}), the optical lines did
not show the presence of double peaks. From our data we are not able to
distinguish if this is due to the presence of circumstellar matter that
smears the double peaked lines into single peaked ones, or due to the
real absence of an accretion disc. It is worth noting at this point that
there is a class of binaries, namely the \object{SW Sex} cataclysmic
variables, in which the presence of an accretion disc does not
correspond to the presence of double peaked emission lines (see e.g.\
\cite{2037}). Finally, the short outburst duration and the temporal
evolution of the X--ray spectrum from the outburst to quiescence is not
typical of a BH system, which is characterized by a two-component
spectrum (see e.g.\ \cite{2034}): a thermal-like ($kT\sim$1 keV) and a
high energy power-law tail that becomes comparatively stronger as the
source weakens, leading to spectral changes not observed in \S\ (however
\object{GS 2023+338} and \object{GRO J0422+32} do not show the soft
component; see \cite{2033}). Also the observed very high power-law index
(see Table~\ref{tab:fits}) is in contrast with a typical BH system
(\cite{2033}. An exception is \object{A0620--00}; \cite{2038}). Although
we cannot firmly exclude the presence of a BH in \S, for these reasons
we consider it unlikely.

Next the NS case: the presence of an accretion disc is unlikely because
of the same considerations given for the BH case. Also the presence of a
hot, optically thin corona around the NS is excluded by the lack of hard
X--ray emission in quiescence (\cite{2039}). While the majority of
X--rays from a BH system probably comes from an accretion disc, in the
case of a NS system the surface of the compact object can also be a
source of X--rays. The $kT_1\sim220$~eV temperature observed in \S\ and
its X--ray luminosity would correspond to an emitting area of
$\sim$1~Km$^2$, too small to be produced at the surface of a NS unless
funnelling of the accretion by the magnetic field  onto a small area is
introduced. This emitting area estimate is however based on the
assumption of blackbody emission from the NS surface. It has been shown
that because of cooling and back warming effects at the surface the
spectrum differs significantly from that of a blackbody (\cite{1862}),
with the effect that the emitting area can be understimated by as much
as 2 orders of magnitudes. Fits with Hydrogen atmosphere spectral models
to three type I bursting NS systems have shown the consistency of the
emitting area with a 10~Km radius NS (\cite{1904}). Because our $kT_1$
is quite similar to that observed in these three systems, we are not
able to exclude that \S\ contains a NS, although this is not supported
by direct evidence, such as type I X--ray bursts or coherent pulsation
(that could have been smeared out by the circumstellar matter).

We finally discuss the possibility that \S/ CI Cam is a binary system
containing a WD. Thermal emission from the WD surface is not able to
explain the observed quiescent X--ray luminosity unless only a very
small fraction of the WD is responsible of the X--ray emission. Indeed a
typical photon energy of $kT_{\rm bb}\sim 2.5\,\alpha^{-0.25}\,\sqrt{\rm
d_{\rm kpc}}$~eV, where $\alpha$ is the fraction of the WD surface that
emits X--rays, is implied. Therefore the most likely process is direct
accretion from the companion star. For a 1~M$_\odot$, $R_x$=$10^9$~cm
WD, the accretion rate is $\sim$10$^{-12}$~d$^2_{\rm kpc}$
M$_\odot$~yr$^{-1}$. For such a low rate we expect that the main cooling
mechanism is not free-free (bremsstrahlung) emission but thermal
cyclotron emission (\cite{864}), with the consequence that a
two-temperature thermal spectrum is expected. Indeed, ROSAT observations
of AM Her systems in low-state show two-temperature spectra, with
$(kT_1, kT_2)$ in qualitative agreement with that observed in \S\
(\cite{2036}). Anyway, a more detailed analysis, beyond the scope of
this work, is needed to quantitatively describe the quiescence \S\
X--ray spectrum.

In the framework of a WD system scenario, the outburst experienced by
\S\ would be associated to a thermonuclear runaway on the surface of a
$\sim1$~M$_\odot$ hot WD. The calculations performed by Iben (1982;
Section~V, page 254) \nocite{1861} show that a peak duration of about
one week or less, depending on the accretion rate, can be achieved, with
a burst recurrence time $\ga$500 yr. Furthermore, if the WD is young
enough to suffer its first thermonuclear instability then it could be
thermally unsettled (\cite{1922}), and therefore have not yet achieved
the long term equilibrium between accretion rate and nuclear burning,
with the possible consequence of a short post-burst phase. Of course,
for a quantitative description of the outburst a detailed application of
the Iben's calculations, beyond the scope of this paper, must be carried
out. From a qualitatively point of view, the onset of the soft (moving)
component observed during TOO2 (\cite{1775}) would be explained in terms
of ejection of the H- and He-rich layers. This would also explain the
reduction of the He~I intensity observed in the quiescent optical
spectrum. The outburst radio emission would be due to relativistic
electrons accelerated in the outward-moving shock (see e.g.\
\cite{185}), while the quiescent (shell-like) radio emission would be
due to the slower motion of the layers.

The low ionization state lines such as Fe~II observed from the optical
spectrum could then arise on the surface of the cool component because
of irradiation by the hot star (see Hoffmaister et~al.\ (1985)
\nocite{1868} and references therein), while the high-excitation
emission lines such as Fe~XXIV could be due either to an extended
nebular shell shrouding the system, or to the reflection of the X--rays
on the optically thick cold matter on the companion's surface.

Finally, we can exclude that the X--rays observed from \S\ during
quiescence come directly from the optical star if it is of spectral type
OB. Indeed, surveys with the Einstein Observatory showed that OB stars
are emitters of soft ($\la$4 keV) X--ray photons (\cite{1889}).
Pallavicini et~al.\ (1981) \nocite{1890} have shown that for OB stars a
relation between X--ray and bolometric luminosity holds: $L_x/L_{\rm
bol}\sim 10^{-7}$. From a typical effective temperature for a B star of
$T_{\rm eff}\ga 22000$~$^\circ$K we infer $L_{\rm bol}\ga 5\times
10^{37}$~erg~s$^{-1}$. With our value of $L_x$ we obtain $L_x/L_{\rm
bol}\sim 10^{-5}$, at least two orders of magnitude greater that that
expected from OB stars. Therefore we conclude that the observed
quiescent X--ray emission cannot be due to a B star.

\section{Summary and conclusions}

We have shown that the quiescent X--ray spectrum observed from \S\ can
be fit with the same model used soon after its outburst, namely a two
temperature bremsstrahlung. While the lower temperature, of the order of
few hundreds eV, did not change between TOO2 and this observation, the
higher temperature decreased considerably. The optical spectrum is very
complex, and it is not possible to determine a spectral class for the
system components. We discussed the nature of the compact object present
in the system, and taking into account the peculiarity of the source,
and the difficulty to compare its properties with those of typical
systems, we conclude that it is unlikely for the \S/CI Cam system to
contain a BH. Both a NS and a WD are possible, but the WD hypothesis is
more appealing because it better fits the observational scenario: a
two-temperature thermal spectrum for the X--ray emission, the lack of
any temporal variability, and the presence of shell like motion observed
in radio, that can be explained in terms of ejection of H- and He-rich
layers during the outburst due to a thermonuclear runaway. This also
fits nicely with the observed reduction of the He~I line strengths when
compared to observations performed before the outburst.

Finally, we can exclude that the X--ray luminosity observed in
quiescence is due to the optical star if it is of spectral type OB.
Moreover, the optical emission cannot be due to re-processing of the
X--rays because $L_x/L_{\rm opt}\sim 10^{-3}$.

\begin{acknowledgements}
We wish to thank Bob Hjellming for letting us know prior publication the
\S\ distance as obtained from his radio observations. We thank the
anonymous referee for helpful comments that greatly improved the paper.
\B\ is a joint Italian and Dutch programme. This research was supported
in part by the Italian Space Agency.
\end{acknowledgements}


\end{document}